\documentclass[10pt]{iopart}
\expandafter\let\csname equation*\endcsname\relax

\expandafter\let\csname endequation*\endcsname\relax

\usepackage{graphicx}%
\usepackage{multirow}%
\usepackage{amsmath,amssymb,amsfonts}%
\usepackage{amsthm}%
\usepackage{mathrsfs}%
\usepackage{xcolor}%
\usepackage{textcomp}%
\usepackage[noadjust,compress]{cite}
\usepackage{lipsum}

\usepackage[colorlinks=true,linkcolor=blue, citecolor=blue, urlcolor=blue]{hyperref}
\begin{document}
\title[Maxwell's equal-area law for Vaidya-Bonner-de Sitter black hole under LIV]{Maxwell's equal-area law for Vaidya-Bonner-de Sitter black hole under Lorentz invariance violation}

\author{Yenshembam Priyobarta Singh$^1$ \footnote{ E-mail: priyoyensh@gmail.com}, Telem Ibungochouba Singh$^{1,*}$ \footnote{$^*$
 E-mail: ibungochouba@rediffmail.com }, 
Sapam Niranjan Singh$^{1}$ \footnote{ E-mail: sapamkhomba@gmail.com}}
\address{$^1$ Department of Mathematics, Manipur University, Canchipur, Imphal 795003, Manipur, India}


\begin{abstract}
In this paper, we investigate the tunneling of fermions with arbitrary spin near the event horizon of non-stationary Vaidya-Bonner-de Sitter black hole (VBdS) in Lorentz invariance violation (LIV). The modified Hawking temperature of VBdS black holes is calculated by using tortoise coordinate transformation, Feynman prescription and WKB approximation. By considering cosmological constant as a thermodynamic pressure in the extended phase space, we construct Maxwell's equal-area law in LIV  and study the phase transitions of VBdS black hole in $P-\tilde{v}$, $P-V$ and $T-S$ plane. The LIV increases the length of liquid-gas coexistence region. The thermodynamic quantities such as entropy, heat capacity, Helmholtz free energy, Internal energy, enthalpy and Gibbs free energy of VBdS black hole are discussed. These quantities tend to increase under LIV. The stability of black hole is also discussed in the presence of LIV.\\


\end{abstract}
\maketitle

\ioptwocol
\section{Introduction}

The LIGO and Virgo observations, as well as the   M87$^*$  black hole shadow captured by Event Horizon Telescope, have proven the existence of black holes \cite{abbott1,abbott2,akiyama1,akiyama2,akiyama3}. Since then,  the study of black holes and their properties become more relevant and fascinating subjects among the researchers.   The Hawking temperature is a key concept in black hole thermodynamics and it has a great significance in understanding the nature of black holes. Hawking   \cite{hawking1,hawking2}  proposed the thermal radiation   for studying the quantum effects near the event horizon of black hole. Several methods for calculating the Hawking temperature of black holes are found in the literatures \cite{gibbons1,damour1,robson1,ovgun1}. The semiclassical tunneling method proposed by Kraus, Parikh and Wilczek  \cite{ kraus1,kraus2,parikh1,parikh2} for deriving Hawking temperature is widely used to study the Hawking temperature for different types of black holes. In these papers, the radial null geodesic method  is used to derive the Hawking temperature in accordance with the semi-classical WKB approximation. Another tunneling method to derive Hawking temperature is the Hamilton-Jacobi  method \cite{angheben1} which is an extension of the complex path analysis of Padmanabhan et al. work \cite{srinivasan1,shankaranarayanan1,shankaranarayanan2}. Many remarkable results were obtained using this technique in \cite{zhang1,zhang2,kerner1,kerner2,rahman1,ibungochouba1}. Ref. \cite{banerji1,banerji2} corrected the tunneling probability after considering the effects of backreaction and self-gravitation to study the Hawking temperature and semi-classical black hole entropy. The effects of the GUP  in the tunneling formalism for Hawking radiation is widely studied in \cite{adler1,chen1,chen2,li1,ovgun2,feng1,vagenas1,ibungochouba2,carr1,priyo1} and calculated the  quantum corrected Hawking temperature.
  
 According to the study of string theory and quantum gravity theory, Lorentz invariance, which is a fundamental principle in physics, is found to be broken down at high energy case \cite{lambiase1,anacleto1,jha1}. The Lorentz dispersion relation is required to be modified in the high energy case and  the magnitude of this correction term  shoud be in Planck scale \cite{magueijo1,amelino1,kruglov1}. Based on the LIV, the Klein-Gordon equation and Dirac equation are required to modify for curved spacetime \cite{chen3,zhang3,sha1,li2}. This leads to the  correction of physical quantities such
as black hole quantum tunneling radiation, temperature,
 entropy and other thermodynamic quantities. Thus the study of  tunneling radiation of bosons and fermions under LIV is a promising area of research.  In the last few years, based on LIV modification,  many researchers studied the corrected tunneling radiation for different types of black holes \cite{zhang4,jin1,sha2,yang5,liu1,liu2,priyo2,onika1,media1}.
  
 Hawking and Page \cite{hawking3} discovered
a phase transition between the Schwarzschild-AdS black hole and the thermal AdS space. Chamblin et al. \cite{chamblin1} investigated the first order phase transition in Reissner-Nordstrom-AdS (RNAdS) black holes and explored the analogy of phase transition to Van der Waals liquid-gas system both in  canonical ensemble and in grand canonical ensemble. Further, treating the cosmological constant as a thermodynamic pressure, $P=-\frac{\Lambda}{8 \pi}$ and its conjugate quantity as a thermodynamical volume, the phase transitions and critical behaviors of RNAdS black hole in extended phase space are  studied in \cite{dolan1,kubiznak1,kubiznak2}. These phase transitions of RNAdS in the extended phase space are similar to the Van der Waals liquid-gas system's phase transitions. Moreover, the critical exponents are the same with those of Van der Waals liquid-gas system. The $P-V$ criticality of black holes in different modified theories of gravitation are extensively studied in \cite{cai1,zhao1,dolan2,majhi3,zou1,jafarzade1,kumar1}. The Van der Waals liquid-gas system analogous to the $P-v$ criticality of charged dynamical (Vaidya) AdS black hole, including the equation of state and critical exponents were studied in \cite{ li3}. Further Van der Waals like liquid-gas phase transition is observed in the $T-S$ plane of black hole \cite{chamblin2}.
  
For Van der Waals liquid-gas system, above the critical temperature $T_c$ the isothermal curve shows a similar behavior to the experimental results but below $T_c$ there exists an oscillating region which violates the condition of stable equilibrium.   
The theoretical prediction and experimental results are reconciled by substituting the oscillating part of the isotherm with a horizontal isobar that satisfies Maxwell's equal-area law. Ref. \cite{kubiznak1,zhao2,belhaji1,li4} studied the Maxwell's equal-area law in $P-v$ plane. Further, the construction of Maxwell's equal-area law is extended in $P-V$ and $T-S$ planes  \cite{spallucci1,spallucci2,zhang5,guo1,sharif1}. However, no research work has been done so far for the construction of Maxwell's equal-area law under LIV modification. Our work is to construct Maxwell's equal-area law under LIV for VBdS black hole in extended phase space.
  
The organization of the paper is as follows: The correction of quantum tunneling radiation of fermions from VBdS black hole induced by LIV is presented in Section 2. In Section 3, we study the Maxwell equal-area law of VBdS black hole in extended phase space and obtain  the positions of phase transition under LIV modification for different conjugate variables $P-v$, $P-V$ and $T-S$. In Section 4, under LIV modification we present the entropy correction and calculate the modified Helmholtz free energy, internal energy and  enthalpy. In Section 5, the stability of the VBdS black hole under LIV is discussed  through Gibbs free energy,  heat capacity and  Hessian matrix. The last section gives the discussions and conclusions.

\section{ Tunneling of VBdS black hole under LIV}  
\renewcommand {\theequation}{\arabic{equation}}
 The metric of the nonstationary VBdS black hole in an advanced Eddington-Finkelstein coordinate $(v, r, \theta, \phi)$ is defined by \cite{bonner1}
\begin{eqnarray}\label{1}
ds^2  = &&-\bigg(1-\frac{2M}{r} + \frac{Q^2}{r^2}-\frac{\Lambda}{3}r^2\bigg)dv^2 + 2 dv ~dr \cr &&+ r^2(d\theta^2 + \sin^2\theta d\phi^2),
\end{eqnarray}
where $v$ indicates the Eddington time and $ M=M(v)$ and $Q=Q(v)$, are the mass and charge of the black hole respectively and $\Lambda$ is the cosmological constant. The four vector electromagnetic potentials $A_{\mu}$  of VBdS black hole is given by $A_{\mu}=(\frac{Q}{r}, 0, 0, 0)$.        The non zero contravariant components of VBdS black hole are given by
\begin{eqnarray} \label{2}
g^{11} &=& \bigg(1-\frac{2M}{r} + \frac{Q^2}{r^2}-\frac{\Lambda}{3}r^2\bigg), g^{10} = g^{01}=1,\cr\cr 
g^{22} &=& \frac{1}{r^2},\,  g^{33} = \frac{1}{r^2\sin^2\theta}.
\end{eqnarray}
Since the space-time represented by Eq. \eqref{1} is spherically
symmetric, the event horizon is necessary a null surface $r_h=r_h(v)$ that satisfies the null hypersurface condition
\begin{eqnarray}\label{3}
g^{ab}\frac{\partial F}{\partial x^a}\frac{\partial F}{\partial x^b}=0,
\end{eqnarray}
with $F(v,r)=0$.
Using Eq. \eref{1} in Eq. \eref{3}, the horizon equation and mass of the VBdS black hole are 
\begin{eqnarray}
1-\frac{2M}{r_h}+\frac{Q^2}{r_h^2}-\frac{\Lambda}{3}r_h^2 -2\dot{r_h}=0
\end{eqnarray} and
\begin{eqnarray}\label{mass}
M=\frac{r_h}{2}+ \frac{Q^2}{2r_h} -r_h\dot{r}_h - \frac{\Lambda  r_h^3}{6} 
\end{eqnarray}
respectively where $\dot{r}_h = \frac{d r_h}{dv}.$
\subsection{\bf Modified form of Hamilton-Jacobi equation:}  
Refs. \cite{magueijo1,ellis1,kruglov2,jacobson1} investigated a relation in the study of string theory and quantum gravity as
\begin{eqnarray}\label{6}
\tilde{P}_0^2 = \tilde{P}^2 + m^2-(\lambda \tilde{P}_0)^i \tilde{P}^2,
\end{eqnarray}
where $\tilde{P}$ and $\tilde{P}_0$ represent momentum and energy of the particle with the static mass $m$. The constant term $\lambda$ is in the magnitude of Planck scale which is determined from the LIV theory from Eq. \eqref{6}. Then, the value of $i$ is unity in the Liouville string theory. A modified form of Dirac equation is determined from Eq. \eqref{6} for $i=2$ \cite{kruglov1}.
The Rarita-Schwinger-Hamilton-Jacobi equation is given by \cite{zhang4}
\begin{align} \label{11}
& g^{\mu\nu}(\partial_\mu \Psi + eA_\mu)(\partial_\nu \Psi + eA_\nu) + m^2 - 2\lambda m (\partial_v \Psi + eA_0)		\nonumber\\ &		g^{0i}\partial_i \Psi
 -\lambda^2[(\partial_v \Psi + eA_0)g^{0j}\partial_j \Psi]^2 = 0,
\end{align} 
where $\mu,\nu = 0,1,2,3 $ and $i,j=1,2,3$. The action of the fermion can be obtained from Eq. \eqref{11} and corresponding modified Hawking temperature of VBdS black hole can be calculated. Eq. \eqref{11} gives a highly accurate dynamic equation due to presence of the term $O(\lambda^2)$ and  the involvement of LIV. Using Eq. \eqref{2} in Eq. \eqref{11} we get
\begin{eqnarray} \label{12}
 &&\frac{\Delta}{r^2}\bigg(\frac{\partial \Psi}{\partial r}\bigg)^2 + \frac{1}{r^2}\bigg(\frac{\partial \Psi}{\partial \theta}\bigg)^2 + \frac{1}{r^2\sin^2\theta}\bigg(\frac{\partial \Psi}{\partial \phi}\bigg)^2 			\cr
&&			+ 2\bigg(\frac{\partial \Psi}{\partial v} + eA_0\bigg) \bigg(\frac{\partial \Psi}{\partial r}\bigg)  -2\lambda m \bigg(\frac{\partial \Psi}{\partial v} + eA_0\bigg)\bigg(\frac{\partial \Psi}{\partial r}\bigg) 
\cr
&&
-\lambda^2\bigg(\frac{\partial \Psi}{\partial v} + eA_0\bigg)^2\bigg(\frac{\partial \Psi}{\partial r}\bigg)^2 = 0,    
\end{eqnarray} where $\Delta=r^2-2Mr + Q^2- \frac{\Lambda}{3}r^4.$
Since the action $\Psi$ involved in the above equation is a function of coordinates $v, r, \theta, \phi,$ the action $\Psi$ can be derived by using the tortoise coordinate transformation. Therefore the tortoise coordinate transformation is defined by
\begin{eqnarray} \label{13}
r_* &=& r+\frac{1}{2\kappa}ln\frac{r-r_h(v)}{r_h(v_0)},\cr
v_* &=& v-v_0,
\end{eqnarray}
where $\kappa$ and $r_h(v)$ are the surface gravity and location of event horizon respectively. $v_0$ is the initial time where the fermions gets out across the event horizon. The tortoise coordinate transformation describes the spacetime geometry outside the event horizon of VBdS black hole. In this case, $r_*$ approaches to negative infinity near the event horizon of VBdS black hole and $r_*$ approaches to positive infinity when tending to infinite point.  Eq. \eqref{13} can be written as
\begin{eqnarray}\label{14}
\frac{\partial}{\partial r}&=&\frac{1+2\kappa(r-r_h)}{2\kappa(r-r_h)}\frac{\partial}{\partial r_*}, \cr\cr \frac{\partial}{\partial v}&=&\frac{\partial}{\partial v_*}-\frac{\dot{r}_h}{2\kappa(r-r_h)}\frac{\partial}{\partial r_*}.
\end{eqnarray}
To study modified Hawking temperature, the action $S$ can be defined as 
\begin{eqnarray}\label{15}
\Psi = R (v_*, r_*)+ X(\theta, \phi),
\end{eqnarray} and let
\begin{eqnarray}\label{16}
\frac{\partial R}{\partial v_*} = \frac{\partial \Psi}{\partial v_*} = -\omega,
\end{eqnarray}
where $\omega$ is energy of the particle. Using Eqs. \eqref{13}- \eqref{16} in Eq. \eqref{12}, we get
\begin{eqnarray}
&&\frac{1}{2\kappa (r-r_h)}\bigg[\frac{ \Delta}{r^2}\Big(1 + 2\kappa (r-r_h)\Big)^2 - 2\dot{r}_h \big(1 + 2\kappa (r-r_h)\big) \cr
&&
+ 2\lambda m \dot{r}_h\big(1 		+ 2\kappa (r-r_h)\big)  -\lambda^2 \bigg(\frac{\partial R}{\partial v_*} + eA_0\bigg)^2  
\cr
&& \big(1 + 2\kappa (r-r_h)\big)^2 \bigg]
\bigg(\frac{\partial R}{\partial r_*}\bigg)^2 + 2\bigg(\frac{\partial R}{\partial v_*} + eA_0\bigg)
 \cr
&&(1-\lambda m) \big(1 + 2\kappa (r-r_h)\big)\bigg(\frac{\partial R}{\partial r_*}\bigg)+ 2\kappa (r-r_h)
 \cr
&&\Big[m^2 + o(\lambda^2)\Big] =0.
\end{eqnarray}
To obtained the first order term of $\lambda$ in the final outcome, multiplying both sides of the above equation by $2\kappa(r-r_h)$ and taking the limit as $r\longrightarrow r_h$, we get
\begin{eqnarray} \label{18}
\bigg(\frac{\partial R}{\partial r_*}\bigg)^2 - 2(1-\lambda m)(\omega - \omega_0)\bigg(\frac{\partial R}{\partial r_*}\bigg) = 0,
\end{eqnarray}
 where $\omega_0 = eQ/r_h$. To derive the surface gravity near the horizon of VBdS black hole, the limiting value of the coefficient  $\left(\frac{\partial R}{\partial r_*}\right)^2 $ is taken as unity
\begin{eqnarray}
&&\lim\limits_{\substack{ r\to r_h \\
 v\to v_0}} \frac{1}{2\kappa (r-r_h)r^2}\bigg[\Delta[1 + 2\kappa (r-r_h)]- 2r^2\dot{r}_h   + \cr
&&  2\lambda m r^2\dot{r}_h -\lambda^2 r^2 (\omega-\omega_0)^2    \left\lbrace 1 + 2\kappa (r-r_h)\right\rbrace \bigg] = 1.
\end{eqnarray} From the above equation, the surface gravity $\kappa$ is calculated as
\begin{eqnarray}
\kappa = && \frac{1}{2(1-2\dot{r}_h)r_h^3}  \cr && \times\bigg[r_h^2 -Q^2 - \Lambda r_h^4 - 2r_h^2\dot{r}_h(1 + 2\lambda m) \bigg].
\end{eqnarray}
Using Eq. \eqref{14} in Eq. \eqref{18}, we get
\begin{eqnarray}\label{21}
\frac{\partial R}{\partial r} = && \frac{1 + 2\kappa (r-r_h)}{2\kappa (r-r_h)}(1-\lambda m) \cr && \Big[(\omega-\omega_0)\pm (\omega-\omega_0)\Big].
\end{eqnarray}
Integrating Eq. \eqref{21} by applying Feynman prescription near the horizon of black hole, we get
\begin{eqnarray}
R_{\pm} = \frac{\pi i}{2\kappa} (1-\lambda m)\Big[(\omega-\omega_0)\pm (\omega-\omega_0)\Big],
\end{eqnarray}
where $R_+$ and $R_-$ are the outgoing and ingoing waves respectively near the event horizon of black hole. The tunneling probability of fermions is calculated near the event horizon of VBdS black hole in accordance with semiclassical approximation as
\begin{eqnarray} \label{23}
\Gamma = exp (-2 Im \,\Psi) && =exp (-2 Im \,R_{\pm})  \cr 
&&= exp \bigg[-\frac{2\pi(\omega-\omega_0)}{\kappa_0}\bigg] \cr\cr
 &&= exp \bigg(-\frac{\omega-\omega_0}{\it T}\bigg),
\end{eqnarray}
where $\kappa_0 = \kappa/(1-\lambda m)$ represents the modified surface gravity of VBdS black hole due to Lorentz invariance theory. Since Eq. \eqref{23} is similar to Boltzaman formula, the  Hawking temperature of the black hole is derived as
\begin{eqnarray}\label{24}
T &&=\frac{\kappa_0}{2\pi}
\cr && = \frac{r_h-M-2r_h\dot{r}_h(1+m \lambda)-\lambda^2(\omega-\omega_0)^2r_h- \frac{2\Lambda}{3}r_h^3}{2\pi[2Mr_h -Q^2+\lambda^2(\omega-\omega_0)^2r_h^2+\frac{\Lambda}{3}r_h^4](1-\lambda m)}.\nonumber\\
\end{eqnarray}
It is noted from Eq. \eqref{24} that the Hawking temperature and tunneling rate of VBdS black hole have been modified due to presence of correction term $\lambda$. Applying binomial expansion for $(1-\lambda m)^{-1}$ and ignoring higher power of $\lambda$, Eq. \eqref{24} can be written as
\begin{eqnarray}\label{temp}
T 
&=&T_h + \frac{\lambda m}{4\pi r_h(1-2\dot{r}_h)}(1-\frac{Q^2}{r_h^2}-\Lambda r_h^2-3\dot{r}_h),
\end{eqnarray}

where $T_h$ is the original Hawking temperature of VBdS black hole in the absence of the LIV theory which is given by
\begin{eqnarray}\label{tempo}
T_h = \frac{1-\frac{Q^2}{r_h^2}-\Lambda r_h^2-2\dot{r}_h}{4\pi r_h(1-2\dot{r}_h)}.
\end{eqnarray}



\begin{figure}[h!]
\centering
\centerline{\includegraphics[width=235pt]{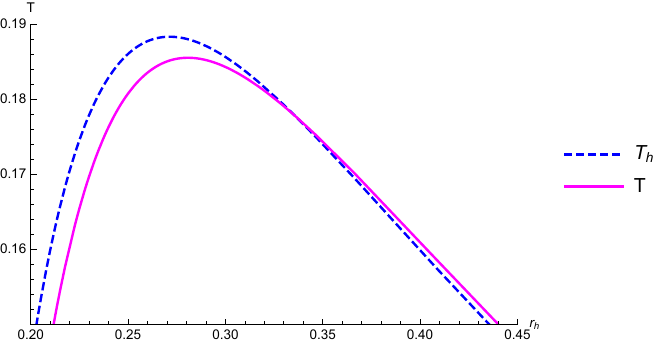}}
\caption{Original and modified Hawking temperature of VBdS black hole versus radius of event horizon $r_h$ for $Q=0.1$, $\lambda=1$, $\Lambda=0.1$, $m=0.1$ and $\dot{r_h}=0.3$. } 
\end{figure}
 In our analysis, we consider the cosmological constant $\Lambda$ as a thermodynamic pressure $P$ 
 \begin{align}
 P=-\dfrac{\Lambda}{8 \pi}.
 \end{align}
 Since  $\Lambda>0$  in de Sitter space, $P$ is negative.  Although, it is more appropriate to take $P$ as a tension than a pressure, we shall continue to refer to it as pressure. Refs. \cite{kubiznak2,dolan2,simovic1,ma1} studied the thermodynamics of de Sitter black holes by treating the positive cosmological constant as  thermodynamic pressure.  Further the corresponding conjugate thermodynamic volume is given as
\begin{align}
V=\dfrac{4}{3} \pi r_{h}^3.
\end{align}


\section{Equal area law of VBdS black hole in extended phase space}
In this section, we will investigate the corresponding Maxwell's equal-area law of VBdS black hole under LIV theory. The equation of state of the VBdS black hole under LIV theory may be obtained from Eq. \eqref{temp} and can be written as $f(T,P,V)=0$. We construct the phase transition of VBdS black hole in $P-\tilde{v}$, $P-V$ and $T-S$ respectively based on Maxwell's equal-area law.

\subsection{Construction of the equal-area law in $P-\tilde{v}$ diagram}
The equation of state for the VBdS black hole under LIV theory is obtained from Eq. \eqref{temp} as
\begin{align}\label{pvr}
P=\dfrac{T(1-2\, \dot{r_h})}{2r_h (1+m \lambda)}+\dfrac{Q^2}{8\pi r_{h}^4}-\dfrac{\Sigma}{8\pi r_h^{2} (1+m \lambda)}.
 \end{align}
 where $\Sigma = 1 + \lambda m -2\dot{r}_h - 3\lambda m \dot{r}_h$.
The equation of state reduces to
\begin{align} \label{pv}
P=\dfrac{T(1-2 \, \dot{r_h})}{\tilde{v} (1+m \lambda)}+\dfrac{2 Q^2}{\pi \tilde{v}^4}-\dfrac{\Sigma}{2 \pi \tilde{v}^2 (1+m \lambda)},
\end{align}
where $\tilde{v}=2 r_h$ is the specific volume. Eq. \eqref{pv}  is used to illustrate $P-\tilde{v}$ curves at constant $Q$ for a given temperature $T$.

\begin{figure}[h!]
\centering
\centerline{\includegraphics[width=235pt]{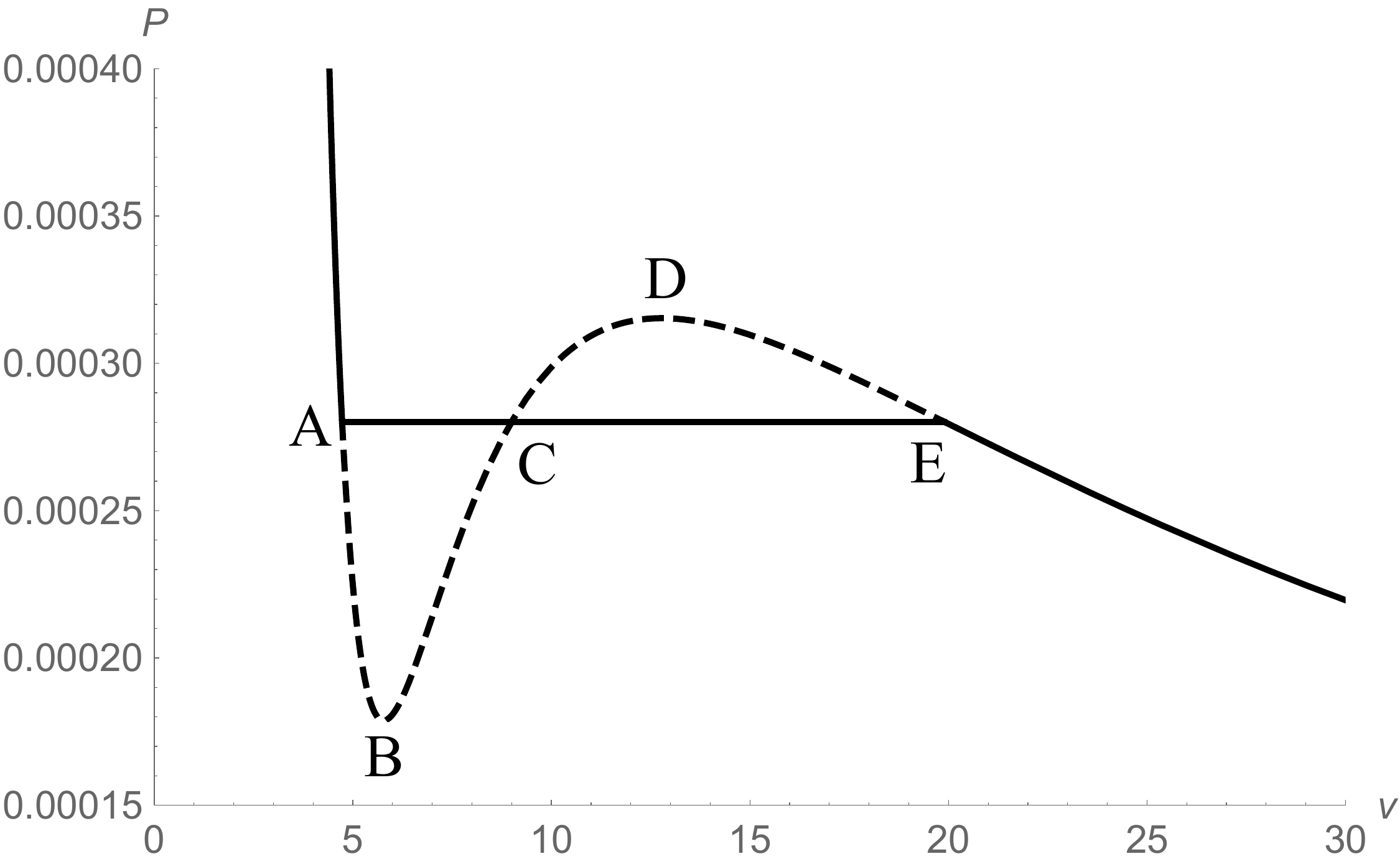}}
\caption{$P-\tilde{v}$ diagram below the critical temperature $T_c$. \textbf{Here we set $m=0.1$, $Q=1$, $\dot{r_h}=0.3$ and $\lambda=0.1$.} } 
\label{fig pv}
\end{figure}
From the Fig. \ref{fig pv}, we observe that there is a phase where one
value of the pressure $P$ corresponds to three different values of $\tilde{v}$ in the isothermal curves  below the critical temperature $T_c$. Experiment results show that there
should be a horizontal isobar in the isotherm to represent
the condensation line where the gas coexists with the liquid.

The thermodynamic system's chemical potential should satisfy
\begin{align}
d\mu=-S dT+V dP.
\end{align}

In the isotherm transformation, the difference of chemical
potential between two states with the pressure $P$ and $P_0$ should be
\begin{align} \label{cp}
\mu-\mu_0=\int_{P_0}^{P} VdP.
\end{align}

From Fig. \ref{fig pv}, it is noted that the black hole is in the ``gas" phase at point E. But the black hole is entirely in the ``liquid" phase at point ``A." Furthermore, the region between A and E may be considered as a coexistence phase.
Since the segment BD defies the equilibrium criteria, the oscillating portion of the curve between A and E cannot be the coexistence line. The chemical potentials are the same at points A and E, which is the thermodynamic condition for phase equilibrium. From Eq. \eqref{cp} we get

\begin{align}
\int_{EDCBA} \tilde{v}~ dP=0,
\end{align}
showing that area ABC is equal to area CDE.

We will find the position of the points A and E for VBdS black hole under LIV and will also discuss the effect caused by Lorentz invariance violation parameter $\lambda$.


The specific volumes at the boundary of the two-phase coexistence area  with a temperature $T_0<T_c$ are $\tilde{v}_1$ and $\tilde{v}_2$ respectively for the VBdS black hole. The corresponding  equal area isobar $P=P_0$, is defined by the event horizon radius $r_h$ and is smaller than the critical pressure $P_c$. Thus, from Maxwell's equal area law, we have

\begin{align}
P_0 \left( \tilde{v}_2-\tilde{v}_1\right)=& \int_{\tilde{v}_1}^{\tilde{v}_{2}} P~d\tilde{v} \nonumber\\ &=\int_{r_1}^{r_2} \left( \dfrac{T_0(1-2 \dot{r_h})}{2 r (1+m \lambda)}+\dfrac{ Q^2}{8\pi r^4}-\dfrac{\Sigma}{8 \pi r^2} \right)
\nonumber\\ &~~~~~~~~~~ \times 2 dr.
\end{align}
Then we can obtain

\begin{align}\label{pv3}
2 P_0 \left(r_2-r_1\right)=&	\dfrac{T_0 \left(1-2 \dot{r_h}\right)}{\left(1+m\lambda\right)} \ln\left( \dfrac{r_2}{r_1}	\right)-		\dfrac{Q^2}{12\pi}\left( \dfrac{1}{r_2^{3}}-\dfrac{1}{r_1^{3}}	\right) 
\nonumber\\
&+\dfrac{\Sigma}{4\pi \left(1+m\lambda\right)} \left( \dfrac{1}{r_2}-\dfrac{1}{r_1}	\right).
\end{align}
From Eq. \eqref{pvr}, we have
\begin{align}\label{pv1}
&P_0=\dfrac{T_0(1-2 \dot{r_h})}{2r_1 (1+m \lambda)}+\dfrac{Q^2}{8\pi r_{1}^4}-\dfrac{\Sigma}{8\pi r_1^{2}}
\end{align}
and
\begin{align}\label{pv2}
& P_0=\dfrac{T_0(1-2 \dot{r_h})}{2r_2 (1+m \lambda)}+\dfrac{Q^2}{8\pi r_{2}^4}-\dfrac{\Sigma}{8\pi r_2^{2}},
\end{align}
where $r_1$ and $r_2$ are the event horizon radii of the $\tilde{v}_1$ and $\tilde{v}_2$ respectively.
 
 From Eqs. \eqref{pv1} and \eqref{pv2} and setting $x=\frac{r_1}{r_2}$,  we  get
\begin{align}\label{pv4}
0= & \dfrac{T_0\left(1-2 \dot{r_h} \right)}{\left(1+m\lambda \right)}+\dfrac{Q^2 \left(1+x\right) \left(1+x^2\right)}{4 \pi x^3 r_{2}^3}
\nonumber\\ &-\dfrac{\Sigma (1+x)}{4 \pi r_2 x \left(1+m \lambda\right)}
\end{align}

and
\begin{align}\label{pv5}
2 P_0= &\dfrac{T_0\left(1-2 \dot{r_h} \right) \left(1+x\right)}{2 r_2 x \left(1+m\lambda \right)}+\dfrac{Q^2 \left(1+x^4\right) }{8 \pi x^4 r_{2}^4}
\nonumber\\ &-\dfrac{\Sigma (1+x^2)}{8 \pi r_2^{2} x^2 \left(1+m \lambda\right)}.
\end{align}
Eq. \eqref{pv3}  can be written as
\begin{align}\label{pv6}
2P_0=&\dfrac{Q^2 (1+x+x^2)}{12 \pi x^3 r_{2}^4}-\dfrac{\Sigma}{4 \pi x r_{2}^2 \left(1+m\lambda \right)}
\nonumber\\ &
-\dfrac{T_0 \ln x \left(1-2 \dot{r_h} \right)}{r_2 \left(1+m \lambda\right) \left(1-x\right)}.
\end{align} 
 From Eqs. \eqref{pv5} and \eqref{pv6}, we find that
 \begin{align}\label{pv7}
& \dfrac{4 \pi r_2 x T_0  \left(1-2 \dot{r_h}\right) \left(1-x^2+2x \ln{x}	\right) -\Sigma \left(1-x\right)^3}{\left(x-1\right) \left(1+m \lambda\right)}
 \nonumber\\ & = \dfrac{Q^2 \left(1-x\right)^2 \left(3x^2+4x+3\right)}{3 r_{2}^2 x^2}.
 \end{align}

Utilizing Eq. \eqref{pv4} in Eq. \eqref{pv7}, we find
\begin{align}\label{pv8}
r_{2}^2=\dfrac{Q^2 \left(1+m \lambda\right)}{3 \Sigma } \times \dfrac{y_1(x)}{y_2(x)},
\end{align}
where
\begin{align*}
& y_1(x)= 4-4x^3 +3 \left(1+x+x^2+x^3\right)\ln{x} \\
\text{and}~~~ &y_2(x)=x^2 \left[ 2-2x+(1+x) \ln{x}\right].
\end{align*}

When $x \rightarrow 1$, we must have $r_1=r_2=r_c$. Therefore from Eq. \eqref{pv8}, we obtain
\begin{align}\label{rc2}
r_c^{2}=& \dfrac{Q^2 \left(1+m \lambda\right)}{3 \Sigma } \lim_{x \to 1} \dfrac{y_1(x)}{y_2(x)}.
\end{align}
By using L'Hopital rule, Eq. \eqref{rc2} becomes
\begin{align}
r_c=\left\lbrace \dfrac{6Q^2\left(1+m \lambda \right)}{\Sigma}\right\rbrace^{\frac{1}{2}}.
\end{align}
It follows that
\begin{align}
& \tilde{v}_c=\dfrac{2 Q\sqrt{6 \left(1+m\lambda\right)} }{\sqrt{\Sigma}}
\end{align}
and
\begin{align}
 & T_c= \dfrac{\Sigma^\frac{3}{2}}{3\sqrt{6} \pi Q \left(1-2\dot{r_h}\right) \sqrt{1+m \lambda}}.
\end{align}
Substituting the value of $r_2$ in Eq. \eqref{pv4} and taking $T_0=\chi T_c$, we can obtain
\begin{align}
2\sqrt{2} \chi x^3  \left( \dfrac{y_1(x)}{y_2(x)}\right)^\frac{3}{2}=9 (1+x) \left[ x^2   \dfrac{y_1(x)}{y_2(x)} -3x^2-3 \right].
\end{align}
\begin{figure}[h!]
\centering
\centerline{\includegraphics[width=235pt]{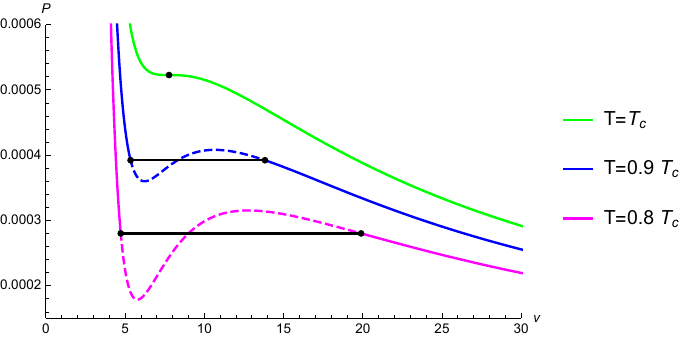}}
\caption{The simulated phase transition (black solid lines) and the boundary of a two phase
coexistence on the base of isobar in the $P-v$ diagram for  VBdS black hole under LIV theory with $m=0.1$, $Q=1$, $\dot{r_h}=0.3$ and $\lambda=0.1$. The temperature of isotherms decreases from top to bottom.} 
\label{fig pvT}
\end{figure}
\begin{figure}[h!]
\centering
\centerline{\includegraphics[width=235pt]{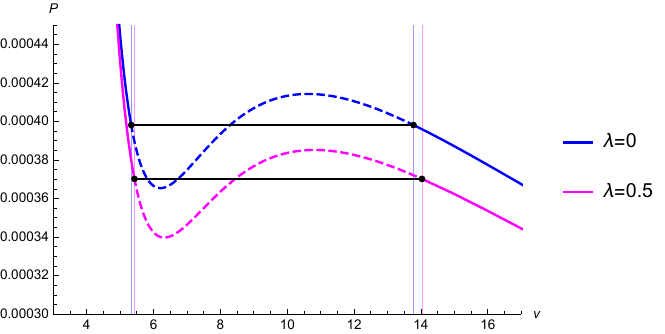}}
\caption{The simulated phase transition (black solid lines) and the boundary of a two phase
coexistence for $T<T_c$ in the $P-\tilde{v}$ diagram for (a) $\lambda=0$ (top) and (b) $\lambda=0.5$ (bottom) with fixed  $m=0.1$, $Q=1$ and  $\dot{r_h}=0.3$. } 
\label{fig pvL}
\end{figure}
 We plot the pressure $P$ in terms of specific volume $\tilde{v}$ for different values of $T$ and $\lambda$ in Figs. \ref{fig pvT} and \ref{fig pvL} respectively and also show simulated phase transition and the boundary of a two-phase coexistence on the base of isotherms in the $P-\tilde{v}$ diagram. From Fig. \ref{fig pvT}, it is noted that when the temperature increases, the isobar in the isotherm becomes shorter. The boundary of the isobar coincides when the temperature reaches its critical value. Moreover, from Fig. \ref{fig pvL}, we observe that under the influence of LIV theory, the phase transition process becomes longer. Further, comparing the original and modified pressure, it is noted that the phase transition occurs at a lower pressure due to LIV theory.
 

\begin{table*}[]
\caption{Numerical solutions for $x$, $r_1$, $r_2$, $\tilde{v}_1$, $\tilde{v}_2$ and $P_0$ for different values of $\lambda$  with $m = 0.1$, $Q=1$ and $\dot{r_h}=0.3$.}
\label{tab pv}
\centering
\begin{tabular}{cccccccc}

\multicolumn{1}{c}{$\lambda$} & \multicolumn{1}{c}{$\chi$} & \multicolumn{1}{c}{$x$} & \multicolumn{1}{c}{$r_1$} & \multicolumn{1}{c}{$r_2$} & \multicolumn{1}{c}{$v_1$} & \multicolumn{1}{c}{$v_2$} & \multicolumn{1}{c}{$P_0$} \\ \hline
\multirow{3}{*}{0}      & 1                      &  1                    &                      3.87298 &                            3.87298                 &   7.74597                    &       7.74597     & 0.000530516           \\
                       &    0.9                   &   0.386969                    &                      2.66568 &      6.88861 &      5.33135           &         13.7772              &                       0.000398128                      \\
                       &    0.8                   &      0.237789                 &                      2.35641 &     9.90968                  &       4.71283                &                      19.8194 &      0.000284384                 \\ \hline
\multirow{3}{*}{0.1}    &  1 & 1 & 3.88744 & 3.88744 &7.77489  &7.77489  & 0.000522667 \\
                      & 0.9 & 0.386969 & 2.67563 & 6.91433 & 5.35126 & 13.8287 & 0.000392237 \\
                       &0.8 & 0.237789 & 2.36521 & 9.94668 & 4.73043 & 19.8934 & 0.000280176                       \\ \hline
\multirow{3}{*}{0.5}      &    1                   &    1                   &                      3.94405 &   3.94405                    &     7.88811                  &                      7.88811 &         0.000493299              \\
                       &        0.9               &       0.386969                &                      2.71459 &          7.01501             &        5.42919               &                      14.03 &              0.000370198         \\
                       &        0.8               &    0.237789                   &                      2.39966 &           10.0915            &     4.79931                  &                      20.1831 &      0.000264433  \\ \hline                
\end{tabular}
\end{table*}

The numerical values of $x$, $r_1$, $r_2$, $\tilde{v}_1$, $\tilde{v}_2$ and $P_0$  for different values of $\lambda$ for VBdS black hole under LIV theory  are displayed in Table \ref{tab pv}. From  Table \ref{tab pv}, we observe that  the values of $x$ decrease as the values of $\chi$ decrease, but $\lambda$ has no effect on it. The values of $r_2$ and $\tilde{v}_2$ increase   with increasing $\lambda$ but decrease with  increasing $\chi$.  However,  the values of $P_0$ decreases with  increasing $\lambda$  but its value increases with increasing $\chi$.

\subsection{Equal-area law in $P-V$ diagram} 
This subsection examines the condition in which the conjugate variables (P, V) occur under Maxwell's equal area law. Further, we will discuss the phase transition of VBdS black hole in  $P-V$ diagram based on Maxwell's equal-area law. On the isotherm with temperature $T_0$ $(T_0< T_c)$ in $P-V$ diagram, there exist two points $(P_0, V_1)$ and $(P_0,V_2)$ satisfying Maxwell's equal area law,

\begin{align}\label{PV1}
P_0(V_2 - V_1)=\int_{V_1}^{V_2} PdV = &\int_{r_1}^{r_2}  \bigg( \dfrac{T_0(1-2 \dot{r_h})}{2 r (1+m \lambda)}+\dfrac{ Q^2}{8\pi r^4}
\nonumber\\ &
 -\dfrac{\Sigma}{8 \pi r^2} \bigg) \times 4 \pi r^2 dr.
\end{align}


%
%
%


 
From Eq. \eqref{PV1}, one can derive the relation
\begin{eqnarray}\label{PV7}
2P_0 = && \frac{3Q^2}{4\pi r_2^4 x(1+x+x^2)} + \frac{3T_0 (1+x)(1-2\dot{r}_h)}{2r_2 (1+\lambda m)(1+x+x^2)} \cr
&&  - \frac{3\Sigma}{4\pi r_2^2(1+x+x^2)(1+\lambda m)}.
\end{eqnarray}
Similarly we can obtain
\begin{eqnarray}\label{PV9}
r_2^2 = \frac{Q^2 (1+4x+x^2)(1+\lambda m)}{\Sigma x^2}.
\end{eqnarray} 


 Taking $T_0 = \chi T_c$, when $0<\chi<1$, we get 
\begin{eqnarray}\label{PV11}
\chi = \frac{3\sqrt{6} x (1+x)}{(1+4x + x^2)^{3/2}}.
\end{eqnarray}

\begin{figure}[h!]
\centering
\centerline{\includegraphics[width=235pt]{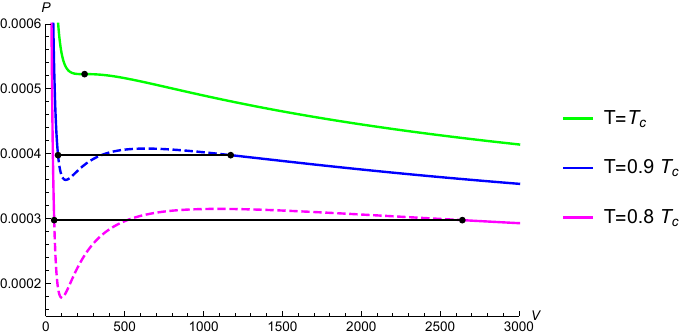}}
\caption{The simulated phase transition (black solid lines) and the boundary of a two
phase coexistence on the base of isobar in the $P-V$ diagram for VBdS black hole under LIV theory with $m = 0.1$, $Q = 1$, $\dot{r_h} = 0.3$ and $\lambda = 0.1$. The temperature of isotherms decrease from top to bottom.} 
\label{fig PVT}
\end{figure}

\begin{figure}[h!]
\centering
\centerline{\includegraphics[width=235pt]{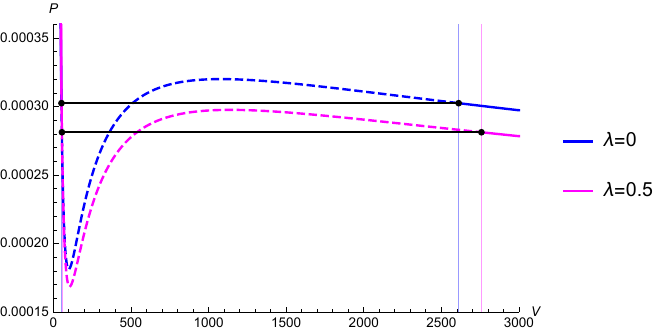}}
\caption{The simulated phase transition (black solid lines) and the boundary of a two
phase coexistence for $T < T_c$ in the $P-V$ diagram for (a) $\lambda = 0$ (top) and (b) $\lambda = 0.5$
(bottom) with fixed $m = 0.1$, $Q = 1$ and $\dot{r_h} = 0.3$. } 
\label{fig PVL}
\end{figure}

The critical state is obtained when $x \rightarrow 1$ i.e $\chi \rightarrow 1$. Using Eqs. \eqref{PV9} and \eqref{PV7}, we can solve $r_2$ and $P_0$ for different values of $\lambda$ and a fixed $\chi$ by deriving a specific value of  $x$ from Eq. \eqref{PV11}. Using the values of $r_2$, we can find the values of $r_1$ and the corresponding values of $V_1$ and $V_2$ can be obtained. To investigate the impact of parameter $\lambda$ on phase transition processes, by fixing the parameters $m=0.1$, $Q=1$, $\dot{r_h}=0.3$ and  $\chi=0.8, 0.9, 1$, the values of $r_1$, $r_2$, $V_1$, $V_2$ and $P_0$ are obtained. The results are shown in Table \ref{tab PV}. From Table \ref{tab PV}, it is shown that $x$ is not related to the LIV parameter $\lambda$. Increasing the values of  $\lambda$, the values of $r_2$ and $v_2$ increase but increasing the values of $\chi$, the values of $r_2$ and $v_2$ decrease. Further, $P_0$ reduces with increasing $\lambda$ implying that the LIV theory lowers the pressure of phase transition. 
We plot $P-V$ daigram for different values of $\chi$ and $\lambda$ in Figs. \ref{fig PVT} and \ref{fig PVL} respectively and show the isobar representing the process of isothermal phase
transition or the two-phase coexistence situation like that of Van der Waals system. Fig. \ref{fig PVT} shows that as the temperature increases the isothermal phase transition process becomes shorter and when the temperature reaches its critical temperature, it turns into a single point. Further, from Fig. \ref{fig PVL} we observe that the isothermal phase transition process becomes longer under the influence of LIV theory.
Moreover, the isobar representing the two phase coexistence occurs at lower pressure under the influence of Lorentz invariance violation theory.

\begin{table*}[h]
\caption{Numerical solutions for $x$, $r_1$, $r_2$, $V_1$, $V_2$ and $P_0$ for different values of $\lambda$  with $m = 0.1$, $Q=1$ and $\dot{r_h}=0.3$.}
\label{tab PV}
\centering
\begin{tabular}{cccccccc}
               $\lambda$   & $\chi$  & $x$  &$r_1$  & $r_2$ & $V_1$ &$V_2$  &$P_0$  \\ \hline
\multirow{3}{*}{0} & 1 & 1 &3.87298  & 3.87298 & 243.347 & 243.347 & 0.000530516 \\
                  & 0.9 & 0.404703 & 2.63752 & 6.51716 & 76.8554 & 1159.48 & 0.000403993 \\
                  & 0.8  & 0.272204 & 2.32535 & 8.5427 & 52.6691 & 2611.4 & 0.000302491 \\ \hline
\multirow{3}{*}{0.1} & 1  & 1 & 3.88744 & 3.88744 & 246.083 & 246.083 & 0.000522667 \\
                  & 0.9 & 0.404703 & 2.64736 & 6.5415 & 77.7194 & 1172.52 & 0.000398015 \\
                  & 0.8 & 0.272204 & 2.3304 & 8.5746 & 53.2613 & 2640.76 & 0.000298015 \\ \hline
\multirow{3}{*}{0.5} & 1 & 1 & 3.94405 & 3.94405 & 256.99 & 256.99 & 0.000493299 \\
                  & 0.9 & 0.404703 & 2.68592 & 6.63675 & 81.1644 & 1224.49 & 0.000375652 \\
                  & 0.8 & 0.272204 & 2.36803 & 8.69946 & 55.6221 & 2757.82 & 0.00028127 \\ \hline
\end{tabular}
\end{table*}


\subsection{Equal-area law in $T-S$ diagram} 
In this section, through Maxwell's equal-area law, we will construct the phase transition in $T-S$ for the VBdS black hole under LIV theory.
The equation of state of Eq. \eqref{pvr} can be written as
\begin{align}\label{ts1}
T=&\dfrac{2P \sqrt{S}(1+\lambda m)}{\sqrt{\pi}(1-2 \dot{r_h})}+\dfrac{\Sigma}{4 \sqrt{\pi S}(1-2\dot{r_h})} 
\nonumber\\ &-\dfrac{Q^2 \sqrt{\pi}(1+\lambda m)}{4 S^\frac{3}{2} (1-2 \dot{r_h})},
\end{align}
where $S=\pi r_{h}^2$ is the Bekenstein-Hawking entropy.

For a given charge $Q$, LIV parameter  $\lambda$ and pressure $P_0< P_c$, the entropy at the boundary of the two-phase coexistence region are $S_1$ and $S_2$ respectively and their corresponding temperature is $T_0$ $(T_0 \leq T_c)$. It is worth mentioning that the temperature depends on the horizon radius $r_h$. From Maxwell's equal area law, we have

\begin{align}\label{ts2}
T_0(S_2-S_1) =& \int_{S_1}^{S_2} T ~dS \nonumber\\
=& \int_{r_1}^{r_2}  \frac{2Pr(1+\lambda m)}{1-2\dot{r}_h} + \frac{\Sigma}{4\pi r_h(1-2\dot{r}_h)} \nonumber\\ &-\frac{Q^2 (1+\lambda m)}{4\pi r_h^3(1-2\dot{r}_h)}  \times 2 \pi r_h dr_h.
\end{align}
Then one can derive
\begin{align}\label{ts3}
2\pi T_0 =& \frac{-Q^2(1+\lambda m)}{ x(1+x)(1-2\dot{r}_h)r_2^3}   +\frac{\Sigma}{(1+x)(1-2\dot{r}_h)r_2}
\nonumber\\ &+\frac{8\pi P_0r_2(1+\lambda m)(1+x+x^2)}{3(1+x)(1-2\dot{r}_h)},
\end{align}
\begin{align}\label{ts4}
T_0 =&  \bigg[\frac{-Q^2(1+\lambda m)}{4\pi r_1^3(1-2\dot{r}_h)}+\frac{2P_0r_1(1+\lambda m)}{1-2\dot{r}_h}+\frac{\Sigma}{4\pi r_1(1-2\dot{r}_h)}\bigg]
\end{align}and
\begin{align}\label{ts5}
T_0 =& \bigg[\frac{-Q^2(1+\lambda m)}{4\pi r_2^3(1-2\dot{r}_h)}+\frac{2P_0r_2(1+\lambda m)}{1-2\dot{r}_h}+\frac{\Sigma}{4\pi r_2(1-2\dot{r}_h)}\bigg]. 
\end{align}
 From Eqs. \eqref{ts4} and \eqref{ts5}, we obtain the following equations
\begin{eqnarray}\label{ts6}
&& \frac{Q^2 (1+\lambda m)(1+x+x^2)}{x^3r_2^3}+8\pi P_0(1+\lambda m)r_2  \cr && -\frac{\Sigma}{x r_2} =0
\end{eqnarray} 
and
\begin{eqnarray}\label{ts7}
8\pi T_0 = & \frac{-Q^2 (1+\lambda m)(1+x^3)}{(1-2\dot{r}_h)x^3 r_2^3}+ \frac{\Sigma (1+x)}{(1-2\dot{r}_h)x r_2 } 
\nonumber\\ &+ \frac{8\pi P_0r_2 (1+\lambda m)(1+x)}{(1-2\dot{r}_h)} .
\end{eqnarray} 
Using Eqs. \eqref{ts3} and \eqref{ts7}, we can derive the relation
\begin{eqnarray} \label{ts8}
8\pi P_0 r_2(1+\lambda m)=&\frac{-3 Q^2(1+3x+x^2)(1+\lambda m)}{r_2^3 x^3} \nonumber\\ &+\frac{3\Sigma}{r_2 x}.
\end{eqnarray}
 Using Eq. \eqref{ts8}, Eq. \eqref{ts6} reduces to
\begin{eqnarray}\label{ts9}
r_2^2 = \frac{Q^2(1+4x+x^2)(1+\lambda m)}{\Sigma \;x^2}.
\end{eqnarray} 
From Eq. \eqref{ts6}, we obtain
\begin{eqnarray}
P_0 = \frac{3 x^2 \Sigma^2}{8\pi Q^2 (1+4x+x^2)^2 (1+\lambda m)^2}.
\end{eqnarray}
The critical radius $r_c =r_1=r_2$ and critical pressure $P_c$ are obtained as follows
\begin{eqnarray}\label{ts10}
r_c =\frac{\sqrt{6}Q(1+ \lambda m)^{\frac{1}{2}}}{\Sigma^{\frac{1}{2}}}
\end{eqnarray} 
and
\begin{eqnarray}\label{ts11}
P_c= \frac{\Sigma^2}{96\pi Q^2 (1+\lambda m)^2}.
\end{eqnarray} 
Taking $P_0 = \chi P_c$, when $0<\chi<1$ we get
\begin{eqnarray}
\chi = \frac{36 x^2}{(1+ 4x + x^2)^2}. 
\end{eqnarray}
\begin{figure}[h!]
\centering
\centerline{\includegraphics[width=235pt]{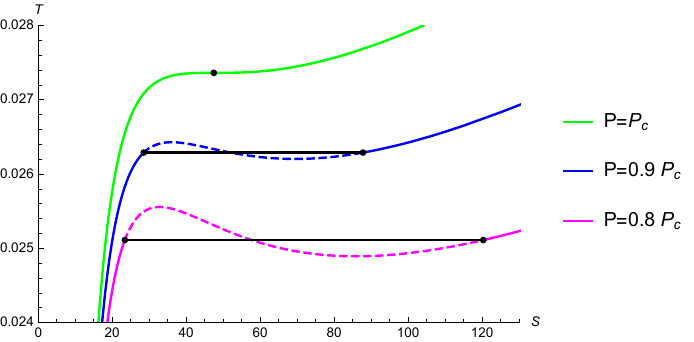}}
\caption{The simulated phase transition (black solid lines) and the boundary of a two phase
coexistence on the base of isobaric in the $T-S$ diagram for  VBdS black hole under LIV theory with $m=0.1$, $Q=1$, $\dot{r_h}=0.3$ and $\lambda=0.1$. The pressure of isotherms decreases from top to bottom. } 
\label{fig ts1}
\end{figure}

\begin{figure}[h!]
\centering
\centerline{\includegraphics[width=235pt]{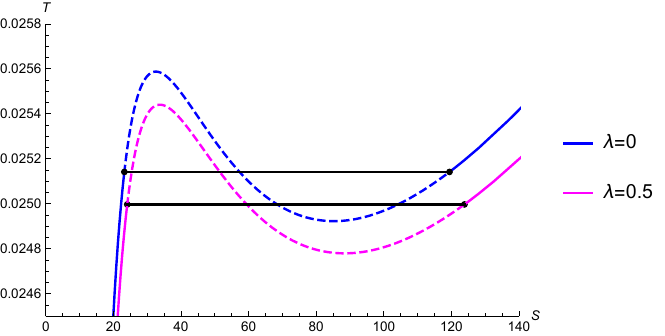}}
\caption{The simulated phase transition (black solid lines) and the boundary of a two
phase coexistence for $P < P_c$ in the $T-S$ diagram for (a) $\lambda = 0$ (top) and (b) $\lambda = 0.5$
(bottom) with fixed $m = 0.1$, $Q = 1$ and $\dot{r_h} = 0.3$. } 
\label{fig ts2}
\end{figure}
We plot $T-S$ diagrams for different values of $\chi$ and $\lambda$ in Figs. \ref{fig ts1} and \ref{fig ts2} respectively and show the isotherms (black solid lines) which represent the simulated
phase transition processes derived from Maxwell's equal area law. From Fig. \ref{fig ts1}, we observe that the isotherm becomes shorter with increasing the pressure and once the pressure reaches its critical point it converges to a
point. For Fig. \ref{fig ts2}, we take $P<P_c$ and plot the $T-S$ diagrams for $\lambda=0$ and 0.5. It is noticed that the phase transition process for $\lambda=0.5$ is longer than that of $\lambda=0$ which implies the LIV theory increases the phase transition process.
\begin{table*}[h]
\caption{Numerical solutions for $x$, $r_1$, $r_2$, $S_1$, $S_2$ and $T_0$ for different values of $\lambda$  with $m = 0.1$, $Q=1$ and $\dot{r_h}=0.3$.}
\label{tab ts}
\centering
\begin{tabular}{cccccccc}
                $\lambda$  & $\chi$ & $x$ & $r_1$ & $r_2$ & $S_1$ & $S_2$ & $T_0$ \\ \hline
\multirow{3}{*}{0} & 1  & 1 & 3.87298 & 3.87298 & 47.1239 & 47.1239 & 0.0273958 \\
                  & 0.9  & 0.569919 & 3.00187 & 5.26719 & 28.3095 & 87.158 & 0.0263212 \\
                  & 0.8 & 0.441089 & 2.7198 & 6.16609 & 23.2393 & 119.445 & 0.0251419 \\ \hline
\multirow{3}{*}{0.1} & 1  & 1 & 3.88744 & 3.88744 & 47.4764 & 47.4764 & 0.0273621 \\
                  & 0.9  & 0.569919 & 3.01307 & 5.28685 & 28.5213 & 87.81 & 0.0262889 \\
                  & 0.8 & 0.441089 & 2.72995 & 6.18911 & 23.4131 & 120.339 & 0.0025111 \\ \hline
\multirow{3}{*}{0.5} & 1 & 1 & 3.94405 & 3.94405 & 48.8692 &48.8692  & 0.0272384 \\
                  & 0.9  & 0.569919 & 3.05695 & 5.36384 & 29.358 & 90.386 & 0.02617 \\
                  & 0.8 & 0.441089 & 2.76971 & 6.27924 & 24.1 & 123.869 & 0.0249975 \\ \hline
\end{tabular}
\end{table*}
We compute the values of $x$, $r_1$, $r_2$, $S_1$, $S_2$ and $T_0$ as $\chi=$ 0.8, 0.9, 1 and $\lambda=$  0, 0.1, 0.5, respectively in order to determine the influence of these parameters on the simulated phase
transition process and the two-phase coexistence region. The tabulated values are displayed in Table \ref{tab ts}. Here, $x$ is unrelated to $\lambda$. $r_2$ and $S_2$ are directly proportional to $\lambda$ but they are inversely proportional to $\chi$. Moreover, $T_0$ is inversely proportional to $\lambda$ and directly proportional to $\chi$. Both $r_2$ and $S_2$ increase with increasing $\lambda$ but decrease with increasing $\chi$.  However, $T_0$ decreases with increasing $\lambda$  but increases with increasing $\chi$.


\section{Thermal Fluctuations}

In this section, we will evaluate the corrected entropy under the influence of LIV theory. To find the corrected entropy, we use the partition function is considered as \cite{pradhan1,das1}

\begin{align}\label{e1}
\tilde{Z}(\beta)= \int_{0}^{\infty} \rho(\tilde{E}) e^{-\beta \tilde{E}} d \tilde{E},
\end{align} 
where $\beta=T^{-1}$. Here $\tilde{E}$ and  $\rho(\tilde{E})$ are the average energy and quantum density of the system respectively.  We apply the inverse Laplace transformation to find the quantum density as
\begin{align}\label{e2}
\rho (\tilde{E})&=\dfrac{1}{2 \pi i} \int_{\beta_0- i \infty}^{\beta_0+ i \infty}  e^{\beta \tilde{E}} \tilde{Z}(\beta) d\beta
\nonumber\\ &=\dfrac{1}{2 \pi i} \int_{\beta_0- i \infty}^{\beta_0+ i \infty}  e^{\tilde{S}(\beta)} d\beta,
\end{align}
where $\tilde{S}=ln (\tilde{Z})+\beta \tilde{E}$ is the corrected entropy for the black hole.

Using  steepest descent method at the saddle point $\beta_0$, the complex integral is calculated such that $\left(\dfrac{\partial \tilde{S}(\beta)}{\partial \beta}\right)_{\beta_0}=0$ and $\dfrac{\partial^2 \tilde{S}}{\partial \beta^2}>0$. Further expanding $\tilde{S}(\beta)$ around the equilibrium $\beta=\beta_0$, we have
%
\begin{align}\label{e3}
\tilde{S}(\beta)=& S+\dfrac{1}{2} \left(\beta-\beta_0	\right)^2 \left(\dfrac{\partial^2 \tilde{S}(\beta)}{\partial \beta^2}\right)_{\beta_0}
\nonumber\\ &+\text{higher order terms},
\end{align}
where $S=\tilde{S}$ is the zero-order entropy and satisfies the relation $\dfrac{\partial S}{\partial \beta}=0$ and $\dfrac{\partial^2 S}{\partial \beta^2}=0$  at $\beta=\beta_0$.
From Eqs. \eqref{e2} and \eqref{e3}, one can derive
\begin{align} \label{e4}
\rho(\tilde{E})=\dfrac{e^S}{2\pi i} \int_{\beta_0-i \infty}^{\beta+i \infty} e^{\frac{1}{2} \left(\beta-\beta_0\right)^2 \frac{\partial^2 \tilde{S}}{\partial \beta^2}} ~ d\beta.
\end{align}

The expression of quantum density can be further simplified as
\begin{align} \label{e5}
\rho(\tilde{E})=\dfrac{e^S}{\sqrt{2\pi}} \left( \dfrac{\partial^2 \tilde{S}}{\partial \beta^2}	\right)^\frac{-1}{2}.
\end{align}

Ignoring the higher order terms and after simplification, we obtain
\begin{align} \label{e6}
\tilde{S}=S-\dfrac{1}{2} ~\text{ln}(S T^2).
\end{align}

%

Using the modified Hawking temperature under the influence of LIV theory ($T$) and the entropy ($S$) in Eq. \eqref{e6}, we obtain the corrected entropy under the influence of LIV theory as
\begin{align}\label{e7}
\tilde{S} =& \pi r_h^2 + ln \bigg[\frac{4\sqrt{\pi}(1-2\dot{r}_h)}{\big(\Sigma - \Lambda r_h^2(1+\lambda m)\big) -Q^2(1+\lambda m)r_h^{-2}}  \bigg].
\end{align}
The corrected entropy in the absence of LIV theory  $S_c$ is calculated using $T=T_h$ as
 \begin{align}\label{e8}
 S_c =& \pi r_h^2 + ln \bigg[\frac{4\sqrt{\pi} r_{h}^2(1-2\dot{r}_h)}{ -Q^2+r_{h}^2 \left(	1-2 \dot{r_h}-\Lambda r_h^2	\right)}  \bigg].
 \end{align}
\begin{figure}[h!]
\centering
\centerline{\includegraphics[width=235pt]{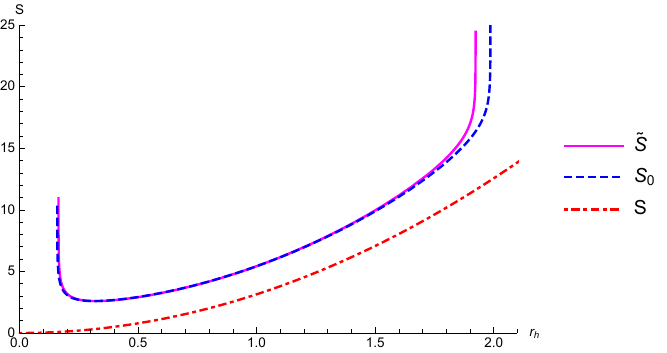}}
\caption{Plot for original entropy, corrected entropy in the absence of LIV theory and corrected entropy under the influence of LIV theory for $Q=0.1$, $\Lambda=0.1$, $m=0.1$ and $\dot{r_h}=0.3$.}
\label{fig entro}
\end{figure}
We plot the original entropy $S$, corrected entropy in the absence of LIV theory $S_c$ and  corrected entropy under the influence of LIV theory $\tilde{S}$ in Fig. \ref{fig entro}. The original entropy ($S$) is monotonically increasing with radius of event horizon $r_h$. But $S_c$ and $\tilde{S}$ initially decrease and later keep on increasing. The LIV theory doesn't have much impact in small black hole  but in the case of larger black hole LIV theory increases the entropy of black holes.

\subsection{Helmholtz free energy}
We analyze the behaviour of Helmholtz free energy of VBdS black hole under the influence of LIV theory. The Helmholtz free energy is given by \cite{akhtar1}
\begin{align} \label{hel1}
F=-\int \tilde{S} \, dT.
\end{align}
Using Eqs. \eqref{temp} and \eqref{e7}, we obtain the expression of Helmholtz free energy under the influence of LIV theory as
\begin{align}\label{hel2}
F=&\frac{1}{12\pi r_h^3(1-2\dot{r}_h)}\bigg[Q^2(2 + 9\pi r_h^2)(1+\lambda m) \nonumber\\ &+ \pi r_h^6(1+\lambda m)\Lambda  + 3r_h^4 \Big\lbrace \big(\pi \Sigma + 2\Lambda(1+\lambda m)\big) \Big\rbrace \cr
&- 3 \Big\lbrace r_h^2\big(\Sigma - \Lambda r_h^2(1+\lambda m) \big)-Q^2(1+\lambda m)  \Big\rbrace  \cr
&\times ln\bigg(\frac{4\sqrt{\pi}(1-2\dot{r}_h)}{\big(\Sigma - \Lambda r_h^2(1+\lambda m)\big) -Q^2(1+\lambda m)r_h^{-2}}\bigg) \bigg].
\end{align}
The Helmholtz free energy without the influence of LIV theory is
\begin{align}\label{hel2o}
F_0=&\frac{1}{12\pi r_h^3(1-2\dot{r}_h)}\bigg[Q^2(2 + 9\pi r_h^2)+ \pi r_h^6 \Lambda  
\cr
&
+ 3r_h^4 \Big\lbrace \big(\pi (1-2\dot{r}_h) + 2\Lambda\big) \Big\rbrace \cr
& -3 \Big\lbrace r_h^2\big(1-2\dot{r}_h - \Lambda r_h^2 \big)-Q^2  \Big\rbrace
\cr
& 
\times  ln\bigg(\frac{4\sqrt{\pi}(1-2\dot{r}_h)}{\big(1-2\dot{r}_h - \Lambda r_h^2\big) -Q^2 \, r_h^{-2}}\bigg) \bigg].
\end{align}

\begin{figure}[h!]
\centering
\centerline{\includegraphics[width=235pt]{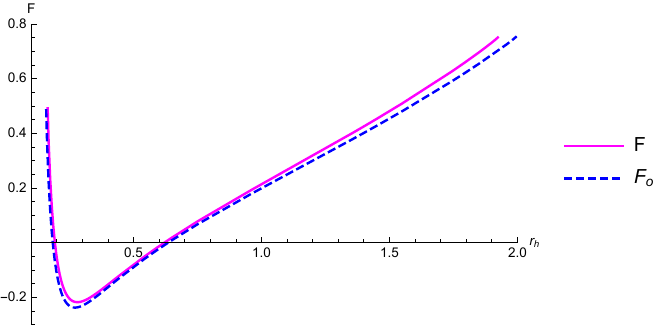}}
\caption{Plot of Helmholtz free energy in the absence of LIV theory and Helmholtz free energy under LIV theory for $Q=0.1$, $\lambda=1$, $\Lambda=0.1$, $m=0.1$ and $\dot{r_h}=0.3$. } 
\label{fig hel}
\end{figure}

 In Fig. \ref{fig hel}, we plot Helmholtz free energy both with and without the influence of LIV theory. Both the Helmholtz free energy show a similar pattern, at first they decrease monotonically upto minimum energy level and later increase with increasing the size of black hole. The LIV theory increases the Helmholtz free energy. In comparison to smaller black holes, larger black holes are  more affected by the LIV theory.
 
\subsection{Internal energy} 
 
The internal energy of VBdS black hole is given by \cite{pourhassan1}
\begin{align} \label{int1}
E= F+T\, \tilde{S}.
\end{align}
Using Eqs. \eqref{temp}, \eqref{e7} and \eqref{hel2}, the internal energy under LIV theory is calculated as
\begin{eqnarray} \label{int2}
E&=& \frac{1}{6\pi \, r_h^3 (1-2\dot{r}_h)}\bigg[Q^2(1+3\pi \, r_h^2)(1+\lambda m) \cr
&+& r_h^4 \Big[3\pi \Big((1 +\lambda m)-\dot{r}_h(1 + 2\lambda m) \Big) + (3-\pi \, r_h^2) \cr
&&(1 + \lambda m)\Lambda \Big] \bigg].
\end{eqnarray}

The internal energy without LIV theory is obtain as
\begin{eqnarray} \label{int2o}
E_0&=& \frac{1}{6\pi \, r_h^3 (1-2\dot{r}_h)}\bigg[Q^2(1+3\pi \, r_h^2) \cr
&&+ r_h^4 \Big[3\pi \left(1 -\dot{r}_h \right)
 + (3-\pi \, r_h^2)\Lambda \Big] \bigg].
\end{eqnarray}
\begin{figure}[h!]
\centering
\centerline{\includegraphics[width=235pt]{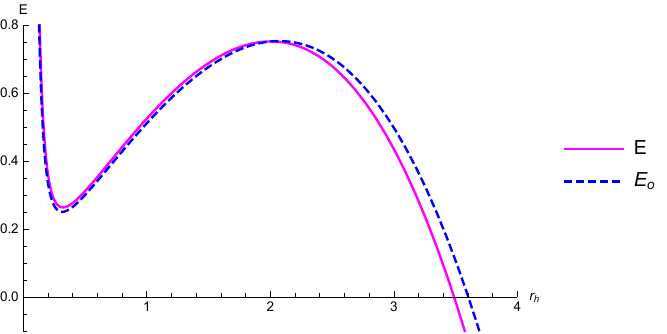}}
\caption{Internal energy of VBdS black hole with and without the influence of LIV theory for $Q=0.1$, $\lambda=1$, $\Lambda=0.1$, $m=0.1$ and $\dot{r_h}=0.3$. } 
\label{fig int}
\end{figure}

We plot Fig. \ref{fig int} for internal energy of VBdS black hole with and without the influence of LIV theory. For small black holes, both the internal energy are positive and show a fluctuation behaviour. However the internal energy becomes negative for large black hole implying the stability of the large black hole. One can see that there exists an event horizon radius $r_h=r_{h}^*$ below which the internal energy under LIV theory is greater than the internal energy without LIV theory  and vice versa for $r_h>r_{h}^*$. The value of $r_h^{*}$ is found to be 2.00781.

\subsection{Enthalpy} 
The enthalpy $H$ of the black hole is given by \cite{akhtar1}
\begin{align}\label{enthalpy1}
H= E + P \,V.
\end{align}

First we derive the pressure from Helmholtz free energy as

\begin{align}\label{enthalpy2}
P=&-\dfrac{d F}{d V}\cr
=&\frac{1}{16\pi^2 r_h^6(1-2\dot{r}_h)}\bigg[r_h^2\Big(\Sigma + \Lambda r_h^2(1+\lambda m)\Big) \cr
& -3Q^2(1+\lambda m) \bigg] \times  \bigg[\pi r_h^2 \cr
& + ln \bigg(\frac{4\sqrt{\pi}(1-2\dot{r}_h)}{\big(\Sigma - \Lambda r_h^2(1+\lambda m)\big) -Q^2(1+\lambda m)r_h^{-2}}\bigg)\bigg]. \cr
\end{align}
Using the Eqs. \eqref{int2} and \eqref{enthalpy2}, we calculate the enthalpy of VBdS black hole under LIV theory  as
\begin{align}
H=& \frac{1}{12\pi r_h^3 (1-2\dot{r}_h)} \bigg[Q^2(2 + 9\pi r^2)(1+\lambda m) \cr &+ r^4\Big[7\pi\Sigma + (6-\pi r^2)(1+\lambda m)\Lambda \Big] \cr
&+ \Big[ r^2\Big\lbrace \Sigma + (1+\lambda m)\Lambda r^2 \Big\rbrace - 3Q^2(1+\lambda m) \Big] \cr
 \times & ln \Big[\frac{4\sqrt{\pi}(1-2\dot{r}_h)}{\Sigma - \Lambda r_h^2(1+\lambda m) -Q^2(1+\lambda m)r_h^{-2}}\Big] \bigg].
\end{align}
The  enthalpy of VBdS black hole in the absence of LIV theory is given by
\begin{align}
H_0=& \frac{1}{12\pi r_h^3 (1-2\dot{r}_h)} \bigg[Q^2(2 + 9\pi r^2)+ r^4\Big[7\pi (1-2 \dot{r_h}) \cr & +\Lambda \left( 6-\pi r^2\right) \Big] + \Big[ r^2 \left( 1-2 \dot{r_h} + \Lambda r^2 \right) - 3Q^2 \Big]   \cr & \times ln \Big[\frac{4\sqrt{\pi}(1-2\dot{r}_h)}{1-2\dot{r_h} - \Lambda r_h^2 -Q^2 r_h^{-2}}\Big] \bigg].
\end{align}

\begin{figure}[h!]
\centering
\centerline{\includegraphics[width=235pt]{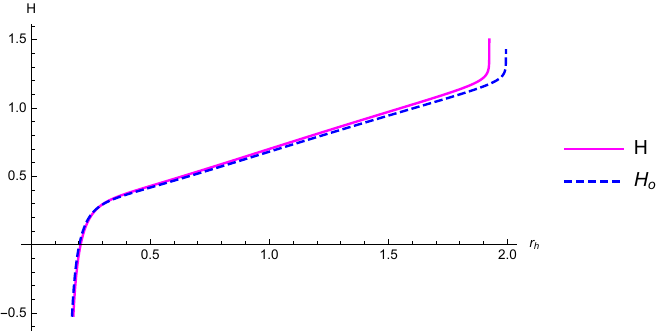}}
\caption{ Enthalpy of VBdS black hole with and without LIV theory for $Q=0.1$, $\lambda=1$, $\Lambda=0.1$, $m=0.1$ and $\dot{r_h}=0.3$. } 
\label{fig enthalpy}
\end{figure}
In Fig. \ref{fig enthalpy}, the enthalpy of VBdS black hole with and without the influence of LIV theory are plotted and illustrated the effect cause by LIV theory. We observe that LIV theory doesn't affect the small black hole but for large black hole LIV theory increases the enthalpy of the black hole.


\section{ Stability of Black hole } 

In this section we discuss the global  and local stability of the black hole. The global stability of the black hole will be analyzed by using the Gibbs free energy. The local stability will also be discussed by evaluating the heat capacity and Hessian matrix.

\subsection{Gibbs free energy}


When the cosmological constant is interpreted as thermodynamic pressure, a new term $V dP$ arises in the first law of black hole thermodynamics. As a consequence, the mass of the black hole is considered as the enthalpy rather than the internal energy \cite{ding1,kastor1,ding2}. Thus the first law of non-rotating charged black hole thermodynamics becomes
\begin{align}
dM=T\,dS+ V\, dP+\Phi \, dQ,
\end{align}
where $\Phi=\frac{Q}{r_h}$ is the electric potential. 
Thus, in canonical ensemble, the Gibbs free energy of the black hole in extended phase space is given by  
\begin{align}\label{gibbs1}
G=M-T S.
\end{align}
The modified Gibbs free energy under LIV theory is obtained by substituting Eqs. \eqref{mass}, \eqref{temp} and \eqref{e7}  in Eq. \eqref{gibbs1} as
%

\begin{align}
G =& \frac{1}{4\pi r_h^3(1-2\dot{r}_h)}\bigg[2\pi r_h^2(1-2\dot{r}_h) \Big(r_h^2 +Q^2-2r_h^2\dot{r}_h \cr & - \frac{\Lambda}{3}r_h^4 \Big)-\Big\lbrace   \Sigma r_{h}^2- (1+\lambda m) \big( Q^2  +\Lambda r_{h}^4 \big) 
 \Big\rbrace \cr &  \Big(\pi r_h^2 + ln \Big[\dfrac{4\sqrt{\pi}r_{h}^2(1-2\dot{r}_h)}{ \Sigma r_{h}^2- (1+\lambda m) \big( Q^2  +\Lambda r_{h}^4 \big) } \Big]\Big)\bigg].
\end{align}
The Gibbs free energy in the absence of LIV theory is obtain as
\begin{align}
G_0 =& \frac{1}{4\pi r_h^3(1-2\dot{r}_h)}\bigg[2\pi r_h^2(1-2\dot{r}_h) \Big(r_h^2 -Q^2-2r_h^2\dot{r}_h \cr & - \frac{\Lambda}{3}r_h^4 \Big)-\Big(   r_{h}^2-2r_{h}^2 \dot{r}_h -Q^2-\Lambda r_{h}^4 \Big) \cr &  \Big(\pi r_h^2 + ln \Big[\dfrac{4\sqrt{\pi}r_{h}^2(1-2\dot{r}_h)}{   r_{h}^2-2r_{h}^2 \dot{r}_h -Q^2-\Lambda r_{h}^4  y } \Big]\Big)\bigg].
\end{align}
\begin{figure}[h!]
\centering
\centerline{\includegraphics[width=235pt]{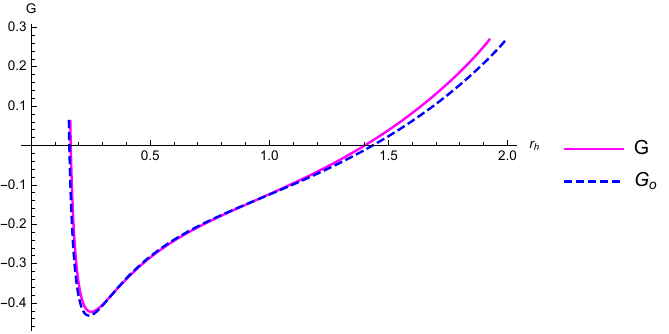}}
\caption{Gibbs free energy of VBdS black hole with or without the influence of LIV theory for $Q=0.1$, $\lambda=1$, $\Lambda=0.1$, $m=0.1$ and $\dot{r_h}=0.3$. } 
\label{fig gibbs}
\end{figure}
The study of Gibbs free energy of black hole provides a vital information about the global stability of black hole.  The preferred phase of the system is the one that minimizes the Gibbs free energy.
Gibbs free energy under LIV modification and in the absence of LIV modification are plotted in Fig. \ref{fig gibbs}. Both the Gibbs free energy decrease for small black hole and  increase with increasing $r_h$. The small black hole has lower Gibbs free energy  and hence it is globally stable. The large black holes have
a higher Gibbs free energy signifying a globally
unstable state. Further, the black holes are more unstable under the influence of LIV theory.



\subsection{Heat Capacity}
The study of the heat capacity of black hole provides a vital information  about its phase transitions and thermodynamic local stability. The phase transition point is the point where the heat capacity either vanishes or diverges. The points where heat capacity vanishes are  the first-type phase transition  whereas the points at which the heat capacity diverges  correspond to the second-type phase transition. Moreover, a stable black hole has a positive heat capacity, whereas an unstable black hole has a negative heat capacity. 

The heat capacity in the absence of LIV theory is calculated using Eqs. \eqref{tempo} and \eqref{e8} as

\begin{align} \label{heat1}
C_0=&T_h\frac{\partial S_c}{\partial T_h}
\cr =& \frac{2\bigg[Q^2(1+\pi r_h^2)- r_{h}^4 \left\lbrace \Lambda+\pi \left(1-2 \dot{r_h}-r_h^2 \Lambda \right)  \right\rbrace    \bigg]}{r_{h}^2 \left(1-2 \dot{r_h}+r_h^2 \Lambda \right) - 3Q^2 }.\cr
\end{align}

Substituting Eqs. \eqref{temp} and \eqref{e7} in Eq. \eqref{heat1}, we find the modified heat capacity  under the influence of LIV theory as
\begin{align}
C =T\frac{\partial S}{\partial T}= &\frac{2}{r_h^2\Big(\Sigma + \Lambda r_h^2(1+\lambda m)\Big) - 3Q^2 (1+\lambda m)}
 \cr &
 \bigg[Q^2(1+\lambda m)(1+\pi r_h^2)- \bigg\lbrace      \Lambda (1+\lambda m)
 \cr & +\pi \Big(\Sigma - \Lambda r_h^2(1+\lambda m) \Big) \bigg\rbrace \bigg].
\end{align}
\begin{figure}[h!]
\centering
\centerline{\includegraphics[width=235pt]{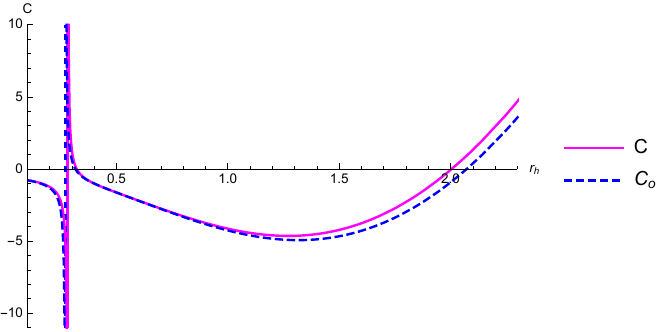}}
\caption{Heat capacity of VBdS black hole with and without LIV theory with respect to  radius of event horizon $r_h$ for $Q=0.1$, $\lambda=1$, $\Lambda=0.1$, $m=0.1$ and $\dot{r_h}=0.3$. }
\label{fig heat} 
\end{figure}
 In order to study the influence of LIV theory on the phase transition and stability of the black hole, we plot $C_0$ and $C$ in Fig. \ref{fig heat}.   For the heat capacity in the absence of LIV theory, the  second-type phase transition point is $r_h=0.271375$ and the first-type phase transition points  are $r_h=0.315466$ and $r_h=2.07203$ for the above set of parameters. However due to LIV theory, the second-type phase transition point is  $r_h=0.280751$ and the first-type phase transition points  are $r_h=0.321609$ and $r_h=2.00468$. We conclude that the positions of phase transitions are affected by LIV theory. It is worth mentioning that the both the heat capacities are positive for large horizon radius implying the stability for large black holes. The stable and unstable range of the heat capacities are displayed in Table \ref{tab stable}.

\begin{table*}[htbp]
\caption{Stable and unstable range of Vaidya-Bonner-de Sitter black hole with and without LIV theory.}
\label{tab stable}
\centering
\begin{tabular}{lll}
                  & Stable range   & Unstable range            \\ \hline
\multirow{2}{*}{\shortstack{Heat capacity \\  without LIV theory $(C_0)$}} & $0.271375<r_h< 0.3$ & $0<r_h< 0.271375$ \\
                  & $2.07203<r_h$ &     $0.315466<r_h< 2.07203$              \\ \hline
\multirow{2}{*}{\shortstack{Heat capacity \\ under LIV theory $(C)$}}		&  $0.280751<r_h< 0.321609$ & $0<r_h<0.280751$  \\
                  & $2.00468<r_h$  &   $0.321609<r_h< 2.00468$               
\end{tabular}
\end{table*}

 \subsection{Hessian matrix}
 
Another approach to check the local stability of black holes is by using Hessian matrix of the Helmholtz free energy. This matrix involve the second order derivatives of Helmholtz free energy with respect to Hawking temperature and chemical potential $(\phi=\frac{\partial M}{\partial Q})$. The Hessian matrix is defined as
\begin{align} \label{eq trace}
\mathcal{\tilde{H}} =\left(\begin{array}{c c}
\mathcal{\tilde{H}}_{aa} & \mathcal{\tilde{H}}_{ab}\\
\mathcal{\tilde{H}}_{ba} & \mathcal{\tilde{H}}_{bb}\\
\end{array}\right)\;\;\; a,b=1,2,
\end{align}
 where $\mathcal{\tilde{H}}_{11} = \frac{\partial ^2 F}{\partial T^2}$, $\mathcal{\tilde{H}}_{12}=\frac{\partial^2 F}{\partial T \, \partial \phi}$, $\mathcal{\tilde{H}}_{21} = \frac{\partial^2F}{\partial \phi \, \partial T}$ and $\mathcal{\tilde{H}}_{22}=\frac{\partial^2 F}{\partial \phi^2}$. The determinant of the Hessian matrix is found to be zero. So, one of the eigenvalue of the matrix  \eqref{eq trace} is zero. The other eigenvalue is given by the trace of the matrix. The trace of the Hessian matrix must be positive for the black hole to be locally stable. The trace of the Hessian matrix is calculated as
 \begin{align}
 \tau=T_r(\mathcal{\tilde{H}}) = \mathcal{\tilde{H}}_{11} + \mathcal{\tilde{H}}_{22},
 \end{align}
 where
 
\begin{align}
H_{11} =& 8\pi r_h^3 (1-2\dot{r}_h)\frac{\Sigma_1}{\Sigma_2},\cr
H_{22} =&\frac{1}{4\pi Q^2 (1-2\dot{r}_h)} \bigg[ 2(1+\lambda m)(3Q^2 + \Lambda r^4) \cr &+\frac{\Theta}{r_h^2\Big(\Sigma - \Lambda r^2(1+\lambda m)\Big)-Q^2(1+\lambda m)} \cr
&\times  ln \bigg\{\frac{4\sqrt{\pi}(1-2\dot{r}_h)}{\big(\Sigma - \Lambda r_h^2(1+\lambda m)\big) -Q^2(1+\lambda m)r_h^{-2}}\bigg\}   \bigg]\cr
\end{align} and the notations $\Sigma_1$, $\Sigma_2$ and $\Theta$ are defined as

\begin{align}
\Sigma_1 =& Q^2(1+\lambda m)(1+\pi \, r_{h}^2)+r_{h}^4\Big[\Lambda(1+\lambda m)(\pi \,r_{h}^2 -1) \cr &-\Sigma \, \pi\Big],\cr
\Sigma_2 =& \Big[3Q^2(1+\lambda m)-\Big(\Sigma + (1+\lambda m)\Lambda r_h^2\Big)r_h^2\Big] \cr &\times \Big[r_h^2\big(\Sigma-\Lambda r_{h}^2(1+\lambda m)\big)-Q^2(1+\lambda m)  \Big],\cr
\Theta =& 2\Big[3Q^4(1+\lambda m)^2 - Q^2 r_{h}^2 (1+\lambda m)\Big\lbrace (1+\pi \, r_{h}^2)\Sigma  \cr & + 2\Lambda \,r_{h}^2(2+\pi \, r_{h}^2)(1+\lambda m)\Big\rbrace + r_{h}^6\Big\lbrace  \Sigma \, \Lambda(1+\lambda m) \cr &(1+\pi \, r_{h}^2) + \pi \, \Sigma^2   + r_{h}^2 \Lambda^2 (1+\lambda m)^2(1-2\pi r_{h}^2) \Big\rbrace  \Big].
\end{align}

\begin{figure}[h!]
\centering
\centerline{\includegraphics[width=235pt]{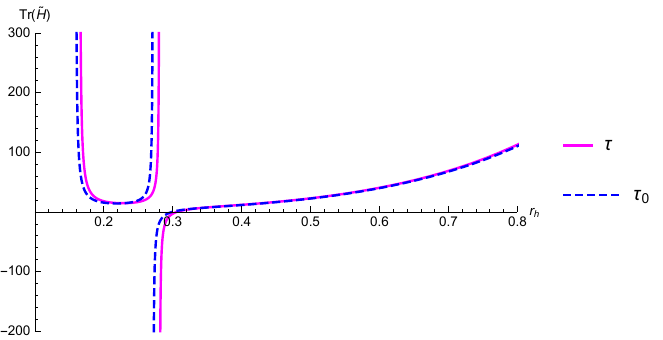}}
\caption{Trace of Hessian matrix with and without the influence of LIV theory for $Q=0.1$, $\lambda=1$, $\Lambda=0.1$, $m=0.1$ and $\dot{r_h}=0.3$. } 
\label{fig hessian}
\end{figure}
 

We plot the trace of the Hessian matrix with respect to event horizon radius in Fig. \ref{fig hessian}.  In the absence of LIV theory, the black holes in the ranges $0.158613<r_h<0.271375$ and $ 0.299026<r_h<\infty$ are stable while under the influence of LIV theory the stable ranges are found to be $0.164393<r_h<0.280751$ and $ 0.305731<r_h< \infty$. The LIV theory affects the stability range of the black hole. Moreover, it is worth mentioning that the large black holes are locally stable.

\section{Conclusion}
In this paper, the quantum tunneling radiation of fermions near the event horizon of VBdS black hole is investigated by using  the Rarita-Schwinger-Hamilton-Jacobi equation under the influence of LIV modification. Based on LIV  modification, the corrected tunneling rate and the Hawking temperature are derived and found to be dependent on the LIV parameter and mass of the particle.
If we take $\lambda=0$, the results are consistent with the paper \cite{li3}. The thermal fluctuations of VBdS black hole under the influence of LIV  modification is also investigated. We use first-order logarithmic corrections to calculate the corrected entropy of VBdS black hole under LIV  modification. Further, the modification due to  LIV  modification in the thermodynamic quantities such as  Helmholtz free energy, internal energy,  enthalpy, Gibbs free energy, and  heat capacity are studied. The results of the paper show that  under the influence of LIV  modification the above thermodynamic quantities tend to increase.  In our graphical analysis, it is observed that the LIV  modification  affects the thermodynamic quantities of large black holes but it does not affect the small black hole. The stability of black hole is investigated by using the Gibbs free energy, heat capacity,  and Hessian matrix.  The local stability range of VBdS black hole under LIV  modification is displayed in Table \ref{tab  stable}. From the above analysis, it is observed that large black holes are locally stable.

Further, the thermodynamic behaviour of VBdS black hole in extended phase space is discussed. By treating the cosmological constant as a thermodynamic pressure, we derive the equation of state under LIV  modification. Similar to Van der Waals liquid-gas
system in the isotherms of the VBdS black holes, we observe a region  where the condition of stable equilibrium violates.  The unphysical
oscillating part in the isotherm should be replaced by an isobar, which represents the liquid-gas coexistence line. For different conjugate variables $P-\tilde{v}$, $P-V$ and $T-S$, we investigate the phase transitions and positions of the boundary of two phase coexistence line using Maxwell's equal-area law and also study the influence of LIV  modification on the phase transition points. In the $P-\tilde{v}$ and $P-V$ planes, the length of the isobar decreases with increasing the temperature. Further, it is noted that the LIV  modification increases the length of phase transition process and the transition of liquid phase to gas phase occurs at a lower pressure due to LIV  modification. Similarly in $T-S$ diagram, we observe that increasing the pressure tends to decrease the length of  isobar. It is noted that the liquid-gas coexistence region in $T-S$ diagram  increases and phase transition occurs at lower temperature under LIV  modification.
  

\section*{Acknowledgements}
 The authors also acknowledge the
anonymous reviewers for valuable suggestions and comments to improve the paper.

\section*{Declaration of competing interest}  The authors declare that they have no known competing financial interests or personal relationships that could have appeared to influence the work reported in this paper

\end{document}